\documentclass[twocolumn]{aastex631}

\usepackage{bm}
\usepackage{xcolor}
\usepackage{amsmath}
\usepackage{times}
\usepackage{accents}

\begin{document}
\title{Lagrangian Perturbation Theory for Biased Tracers:
Significance of the Number Conservation}
\shortauthors{Espenshade \& Yoo}
\author[0000-0002-9293-311X]{Peter Espenshade}
\affiliation{Center for Theoretical Astrophysics and Cosmology, Department of Astrophysics, University of Zürich, Winterthurerstrasse 190, CH-8057, Zürich, Switzerland}
\author[0000-0003-4988-8787]{Jaiyul Yoo}
\affiliation{Center for Theoretical Astrophysics and Cosmology, Department of Astrophysics, University of Zürich, Winterthurerstrasse 190, CH-8057, Zürich, Switzerland}
\affiliation{Department of Physics, University of Zürich, Winterthurerstrasse 190, CH-8057, Zürich, Switzerland}
\email{peter.espenshade@uzh.ch~,~~~~~~jaiyul.yoo@uzh.ch}

\begin{abstract}
\noindent
    The Lagrangian perturbation theory  provides a simple yet powerful way of computing the nonlinear matter power spectrum, and it has been applied to biased tracers such as halos and galaxies. The number conservation of matter particles allows a simple relation between the fluctuations at the initial and the late times, which is essential in deriving the exact expression for the nonlinear matter power spectrum. Here we investigate the significance of the number
conservation in the Lagrangian perturbation theory for biased tracers. 
We use $N$-body simulations to test the significance of number conservation
by tracing dark matter halo samples in time. 
For the mass bin sample $\Delta \log M_h~(h^{-1}M_{\odot})= 0.5$ at~$z\simeq3$,
the theoretical predictions for the 
halos overestimates the power spectrum at $z=0$
 by a factor of three,  while the simulation results match the
theoretical predictions if 
the number conservation of halos is imposed in the simulations 
throughout the evolution. Starting with a halo sample at~$z=0$
as another test, we
trace back in time the particles that belong to the halos at~$z=0$ and use
their center-of-mass positions as halo positions at~$z>0$. The halo
power spectra at~$z>0$ from the simulations agree with the theoretical
predictions of the Lagrangian perturbation theory.
 This numerical experiment proves that the number conservation is crucial in the
Lagrangian perturbation theory predictions.
We discuss the implications for various applications of the 
Lagrangian perturbation theory for biased tracers.
\end{abstract}

\section{Introduction}
The Lagrangian perturbation theory offers a framework to trace the matter distribution from early time to present days (\citealt{Zel'dovich_1970}; see also \citealt{Buchert_1992, Bouchet_1992, Buchert_1993, Buchert_1994_MNRAS, Bouchet_1995, Catelan_1995, BECOET02}). Following the pioneering work by \cite{Crocce_2006_PRD, Matsubara_2008_v77}, it has become an indispensable tool in modeling the bulk motion and the subsequent resummation of the perturbative contributions. More importantly, the Lagrangian perturbation theory provides a simple, but at the same time very powerful way of computing the nonlinear matter power spectrum \citep{Schneider_1995, Taylor_1996, Matsubara_2008_v77}:
\begin{eqnarray}\label{eqn:LPT_theoryPower}
    P_m(k, t) = \int d^3 q ~e^{-i\bm k \cdot \bm q} ~\langle e^{-i \bm k \cdot \Delta \bm\Psi} \rangle~,
\end{eqnarray}
where we have ignored the Dirac delta function at $k=0$,
$\Delta\bm \Psi \equiv \bm \Psi(\bm q_1, t) - \bm \Psi(\bm q_2, t)$, $\bm q \equiv \bm q_1 - \bm q_2$, and $\bm \Psi$ is the displacement field from the Lagrangian position~$\bm q$ at initial time~$t_i$ to the Eulerian position at late time~$t$
\begin{eqnarray}\label{eqn:displacement}
    \bm x (\bm q, t) = \bm q + \bm \Psi (\bm q, t)~.
\end{eqnarray}
These are comoving coordinates (not physical coordinates).
  The expression for the nonlinear matter power spectrum in Equation~(\ref{eqn:LPT_theoryPower}) is fully exact
(see, e.g., \citealt{TACO17,PIETR18,MCVL18,RAFRHA21}).
Hence the task of computing the nonlinear matter power spectrum in practice boils down to
computing the multi-point correlation functions for the displacement field in Fourier space up to a desired order in perturbations \citep{Taylor_1996, Matsubara_2008_v77, Matsubara_2008_v78, Carlson2013,VLSEBA15}.

The critical element in the derivation is the conservation of matter particles
\begin{eqnarray}\label{eqn:numberContinuityMatter}
    n_m(\bm x,t) ~dV = n_m (\bm q,t_i) ~dV_i~,
\end{eqnarray}
where $dV=a^3(t)~d^3x$ and $dV_i=a^3(t_i)~d^3q$ are the physical volumes at times~$t$ and~$t_i$ with the corresponding scale factors~$a(t)$ and~$a(t_i)$ occupied by matter particles, subject to the relation in Equation~(\ref{eqn:displacement}).
  Since the background physical number density decreases~$\bar n_m \propto 1/a^3$  (or the comoving number
density is constant in time),
the conservation equation~(\ref{eqn:numberContinuityMatter}) becomes
\begin{eqnarray}\label{eqn:matterContinuity}
    \left[1+\delta_m(\bm x,t)\right]~d^3x =\left[1+\delta_m(\bm q,t_i)\right]~d^3q~,
\end{eqnarray}
and accounting for the change in the comoving volume factor,
 we obtain
\begin{eqnarray}\label{eqn:jacobian}
1+\delta_m\big[\bm{x}(\bm{q},t),t\big]
=\int d^3q'~\delta^D\big[\bm{x}(\bm{q},t)-\bm{q}'-\bm{\Psi}(\bm{q}',t)\big]~,
\end{eqnarray}
where the fluctuation $\delta_m(\bm{q},t_i)$ at $t_i$ is assumed zero 
by considering only the growing mode and 
setting the initial time $t_i$ to be sufficiently early.
The matter density fluctuation can then be written in Fourier space as
\begin{eqnarray}\label{eqn:matterFourier}
    \delta_m(\bm k, t) = \int d^3 q ~e^{-i \bm k \cdot (\bm q + \bm \Psi)}~,    
\end{eqnarray}
which then allows one to write the exact expression for the matter power spectrum in Equation~(\ref{eqn:LPT_theoryPower}).
Again, the $k=0$ mode is ignored in Equation~\eqref{eqn:matterFourier}.

The Lagrangian perturbation theory has been applied to biased tracers (or halos) \citep{Matsubara_2008_v78, Padmanabhan_2009, Matsubara_2011, Carlson2013, WHITE14}, and the only change in the above equations is that we cannot set the fluctuation~$\delta_h(\bm q, t_i)$ zero at~$t_i$:
\begin{eqnarray}\label{eqn:haloContinuity}
    \left[1+\delta_h(\bm x,t)\right]~d^3x =\left[1+\delta_h(\bm q,t_i)\right]~d^3q~,
\end{eqnarray}
as the biased tracers form in more biased density peaks at early time, while the matter fluctuations are more homogeneous.
However, the key and often understated 
ingredient about the application of the Lagrangian perturbation theory 
to biased tracers is the same number conservation, 
such that the background number density of the biased tracers evolves as
the matter particles, such 
that the background physical
number density of the biased tracers evolves as the matter particles,
\begin{eqnarray}\label{eqn:haloNumberContinuity}
\bar n_h \propto 1/a^3 \propto \bar n_m~,
\end{eqnarray}
or again the comoving number density of the biased tracers is constant
in time.
In fact, dark matter halos as the simplest biased tracer evolve through regular mergers with other dark matter halos and continuous accretion of smooth matter components, invalidating Equation~\eqref{eqn:haloNumberContinuity}.
Halo mergers account
for~$\approx 60\%$ of the total halo mass growth for halo masses spanning the range~$10^9 M_{\odot}$ to~$10^{14} M_{\odot}$ and over the redshift range~$1<z<3$ \citep{MOWH96, Sheth_Tormen_1999, Tinker_2008, Tinker_2010}.
Large-scale galaxy surveys also showed that
the redshift evolution of the mean number density of various galaxy 
samples is different from the matter component with a constant comoving number
density in time or redshift
(see, e.g.,
\citealt{Eisenstein_2001, White_2011, Brammer_2011, Conselice_2016}).
Without the number conservation in Equation~\eqref{eqn:haloNumberContinuity}
and hence Equation~(\ref{eqn:haloContinuity}), one cannot obtain the powerful Equation~(\ref{eqn:haloPerturbation}) for biased tracers, like Equations~(\ref{eqn:LPT_theoryPower}) and~(\ref{eqn:matterFourier}) for matter.

In general, this violation of the number conservation for dark matter
halos poses no problem in the Lagrangian perturbation theory
for biased tracers, as the halo number density~$n_h(\bm q,t_i)$
at early time can represent not the physical number density
of those halos identified at~$t_i$, but the spatial distribution of
the dark matter particles at~$t_i$ that form the halos later in the Eulerian
frame, despite the explicit notation~$n_h(\bm q,t_i)$. Hence the
number of halos remains unchanged throughout the evolution, while those
halos at~$t_i$ cannot be physically identified as distinct halos.
For dark matter particles, in contrast, 
Equation~\eqref{eqn:numberContinuityMatter} represents literally
the number conservation throughout the evolution. In practice,
the Lagrangian perturbation theory for biased tracers is 
applied to the halo samples in the Eulerian frame with 
marginalization of unknown bias parameters, but 
Equation~\eqref{eqn:haloContinuity} remains critical in applying the
Lagrangian perturbation theory for biased tracers.

Here we use dark matter halos in numerical simulations
 to quantify how significant the number conservation is in the 
Lagrangian perturbation theory predictions for biased tracers 
in terms of describing the evolution of dark matter halos and its
clustering. In particular, we first identify dark matter halos 
at high redshift to extract the Lagrangian bias parameters from the simulations,
and we trace the physical evolution of those halos in the simulations
to compute the Eulerian bias parameters at redshift zero.
The Eulerian bias parameters from the simulations are then compared
to the Lagrangian perturbation theory predictions based on the Lagrangian
bias parameters from the simulations.
We find that the Lagrangian perturbation theory predictions match
the numerical simulations, only if the halos number conservation is satisfied.
We discuss the implication of our findings for the 
applications of the Lagrangian perturbation theory in practice
in Section~4.

\section{Biased Tracers}\label{sec:methods}
\subsection{Lagrangian Perturbation Theory for Biased Tracers}
Given the concise and powerful expression for the exact nonlinear matter power spectrum in the Lagrangian perturbation theory, the formalism has been applied to biased tracers \citep{Matsubara_2008_v78, Padmanabhan_2009, Matsubara_2011, Carlson2013, WHITE14}. 
With the number conservation for biased tracers, Equations~(\ref{eqn:matterContinuity}) and~(\ref{eqn:haloContinuity}) can be used in the Lagrangian perturbation theory to model the fluctuation~$\delta_h$ for biased tracers (or halos) at late time
\begin{eqnarray}\label{eqn:perturbationRelation}  
1 + \delta_h[\bm{x}(\bm{q},t),t] = \left[1 + \delta_h(\bm{q},t_i)\right] 
\Big\{1 + \delta_m [\bm{x}(\bm{q},t),t]\Big\}~, 
\end{eqnarray}
where we used Equation~(\ref{eqn:jacobian}) for the Jacobian.

Expanding the relation in Equation~(\ref{eqn:perturbationRelation}) to the linear order in perturbations, we obtain the well-known relation of the Eulerian bias to the Lagrangian bias factor
\begin{eqnarray}\label{eqn:linearBias}
    b_1 = 1 + b_1^\text{L} ~,
\end{eqnarray}
where we defined the bias parameters as
\begin{eqnarray}
    \delta_h(\bm x, t) = b_1 ~\delta_m(\bm x, t)
    ~, \quad
    \delta_h(\bm q, t_i) = b_1^\text{L} ~\delta_m^{(1)}(\bm q, t) ~,
\end{eqnarray}
in terms of the nonlinear matter fluctuation~$\delta_m$ for the Eulerian bias and the linear matter fluctuation~$\delta_m^{(1)}$ extrapolated to late times for the Lagrangian bias. These bias parameters are the key predictions in the Lagrangian perturbation theory that will be tested against numerical simulations. The bias parameters can be further expanded to include higher order corrections, the tidal tensor bias \citep{Szalay_1988, Fry_1993, Catelan_2000, Ma_2000, Baldauf_2012}, as in the effective field theory approach \citep{Mirbabayi_2015, Senatore_2015, VLWHAV15,IVSIZA20}. Here we restrict our investigation to the linear order in perturbations and defer further in-depth investigations of higher order corrections to future work.

\subsection{Numerical Modeling of Biased Tracers}
Here we present our numerical methods to track the evolution of biased tracers from early time to late time. We ran ten dark matter-only PKDGRAV3 \citep{pkdgrav3} simulations with $2048^3$ particles in an $800$~$h^{-1}$Mpc periodic box and $51$~$h^{-1}$kpc softening length yielding a $5.2 \times10^9 ~h^{-1}M_{\odot}$ particle mass. We supplied PKDGRAV3 with the transfer function computed using the Boltzmann solver CLASS \citep{class}, and we used the second-order Lagrangian perturbation theory initial conditions at starting redshift~$z=49.$ We adopt the $\Lambda$CDM parameters: dark matter density~$\Omega_{\text{DM}}=0.27$, baryon density~$\Omega_{b}=0.049$, Hubble constant~$h = 0.67$, scalar spectral index~$n_{s}=0.97$, and primordial power spectrum amplitude~$\ln(10^{10}A_s) = 3.0$, consistent with the Planck results \citep{planck18}.

We use our simulation data to identify dark matter halos as biased tracers by using the friends-of-friends algorithm \citep{Davis_1985} implemented in nbodykit \citep{Hand_2018} with linking length~$b=0.2$. We require that each halo be composed of at least~$100$ particles (or the minimum mass). The halo catalogs are created at each redshift, starting from $z=2.9$ until $z=0$, in units of $\Delta z \sim 0.5$, and each dark matter particle is tagged with a unique identification number, such that we can trace which halo individual particles belong to and thereby how halos at one redshift evolve to the next redshift.

In particular, we focus on a halo sample at early time~$z_\star=2.9$ 
and study how this halo sample evolves forward in time to $z=0$. Note, however, that $z_\star=2.9$ is not the initial time~$t_i$ in the Lagrangian space, but an intermediate time to infer the bias evolution at $z=0$.
We do not consider higher redshift because we need a large number of halos to ensure that their power spectra are not dominated by shot noise. 
At~$z_\star$, we consider a halo sample of
mass $6.6\times10^{11}\leq M(h^{-1}M_\odot)<2.1\times10^{12}$ 
that contains 60\% of halos
with the remaining 30\% halos below and 10\% halos above the halo mass bin,
according to the halo mass function \citep{Tinker_2010}. 
At $z=0$, we consider a halo sample of mass 
$5.0\times10^{13}\leq M(h^{-1}M_\odot)<5.0\times10^{14}$, of which the 
bias parameter is about two, representing  massive galaxies.

To quantify the significance of the number conservation
in the Lagrangian perturbation theory predictions for biased tracers, we compute the power spectra~$P_h$ of the halo samples at $z=0$ (or $z_\star$)
from simulations and compare to the predictions based on the bias parameter obtained from the simulations at $z_\star$ (or $z=0$).
The
Eulerian bias parameter~$b_1$ at $z=0$ is related to the Eulerian bias 
parameter~$b_1$ at~$z_\star$ as \citep{FRY96,TEPE98,DEJESC18}
\begin{eqnarray}\label{eqn:eulerianBiases}
    b_1(z=0) = 1 + \left[b_1(z_\star)-1 \right] \frac{D(z_\star)}{D(z=0)}~,
\end{eqnarray}
where $z_\star=2.9$, and~$D$ is the linear growth function. This equation is derived from Equation~(\ref{eqn:perturbationRelation}) at linear order in perturbations, bypassing the need to compute~$b_1^L$ at the initial time~$t_i$.
We compute the bias parameters from numerical simulations by using $b_1(z) = \sqrt{P_h(z)/P_m(z)}$ over $k \leq 0.03~h~\rm{Mpc}^{-1}$.

\subsection{How Do Halos Evolve through Mergers?}  
Dark matter halos evolve via numerous mergers with other halos and continuous accretion of mass, such that unlike dark matter particles the identities of halos are rather difficult to track through their evolution. For example, the standard method (or the standard merger tree) finds the most massive progenitor of a given halo from the previous output time in a numerical simulation and assigns its identity to the given halo \citep{Benson_2000, Springel_2001, Croton_2006, Kauffmann_1999}. Consequently, when two halos merge into one, a lower-mass halo loses its identity in the standard merger tree. The reality is much more complicated, but this simple scheme makes physical sense when the mass ratio is large.

The critical element
in the Lagrangian perturbation theory for biased tracers is 
that the number of halos is conserved. In the standard approach, the number
conservation of halos is achieved at high redshift~$z$ by considering
dark matter particles that belong to a halo at~$z=0$ as a distinct halo at~$z$,
no matter how much they are spread in space or physically unbound. Here
we first consider different but more physically motivated ways to achieve the
halo number conservation. We 
construct a \textit{number-conserving} merger tree, in which we start with a sample of halos at~$z_\star$ and keep their identities throughout the evolution, even if they merge into a halo of larger mass, no matter how large it is. In case of a merger of two halos with identities we keep, we consider the merged halo as two individual halos at the same position (double counting). In addition, we consider a slight variation of the number-conserving merger tree and construct a \textit{nearly number-conserving} merger tree by not double counting in the number-conserving merger tree.

In short, given a halo sample at~$z_\star$, we construct three different halo samples at $z=0$ by following the halos according to three different identification schemes. Compared to the standard approach, halo samples according to the number-conserving merger tree are more physically traced in time, though once merged into a bigger halo, they cannot be identified as physically distinct halos either. The number of halos, however, remains the same throughout the evolution.
In contrast, a halo sample at $z=0$ obtained by
following the standard merger tree is significantly reduced in number, as
many halos in the sample are incorporated into other larger halos by $z=0$. The crucial difference in three
different halo samples is the validity of the number conservation.

Furthermore, we also test the significance of the
 number conservation in the standard approach to the Lagrangian perturbation
theory
by identifying dark matter particles in the halo sample at $z=0$ and tracing
those particles backward in time. While those particles at $z>0$ are not
physically bound, we treat those particles as if they form a distinct
halo and we use their center of mass as the halo position at $z>0$ in measuring
their power spectrum. In this way, the halo number is also
conserved throughout the evolution backward in time.

\begin{figure}[t]
		{\centering	\includegraphics[width=\linewidth]{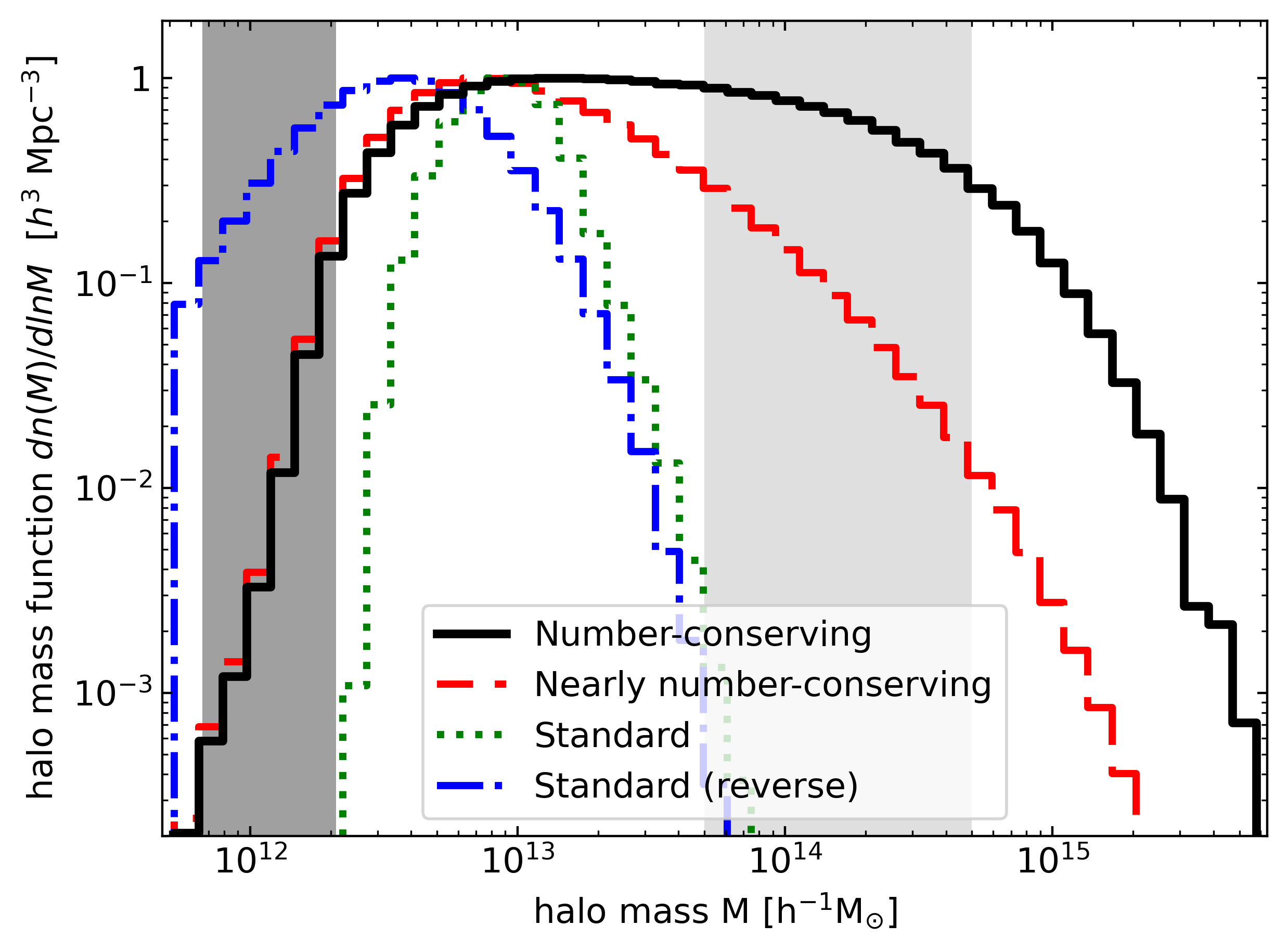}
		\caption{Evolution of halos from a mass-bin sample. Halos in 
the mass-bins sample (dark gray band) at $z_\star=2.9$ are followed forward in time. The curves are normalized to unity to facilitate the comparison. The standard merger tree (dotted) at $z=0$ shows that halos at~$z_\star$ grow in mass, but most~(89\%) of them disappeared by becoming part of larger mass halos via merger. In contrast, if we keep tracking halos at $z_\star$ even when they merge with larger halos, or they merge with each other, the number-conserving merger tree (solid; see text) at $z=0$ shows that the halos in the narrow mass bin (dark gray band) at~$z_\star$ are spread over a large range of mass in halos. The nearly-number-conserving merger tree (dashed; see text) shows the change in the halo distribution from the solid curve if we prevent double counting. In the same way, the standard merger tree can be used to trace back the progenitors at~$z_\star$ (dot-dashed) from the sample at $z=0$ (light gray band).}
		\label{fig:haloMassFunction}
		}
\end{figure}

\section{Comparison to Numerical Simulations}\label{sec:results} 
\begin{figure}[t]
		{\centering	\includegraphics[width=\linewidth]{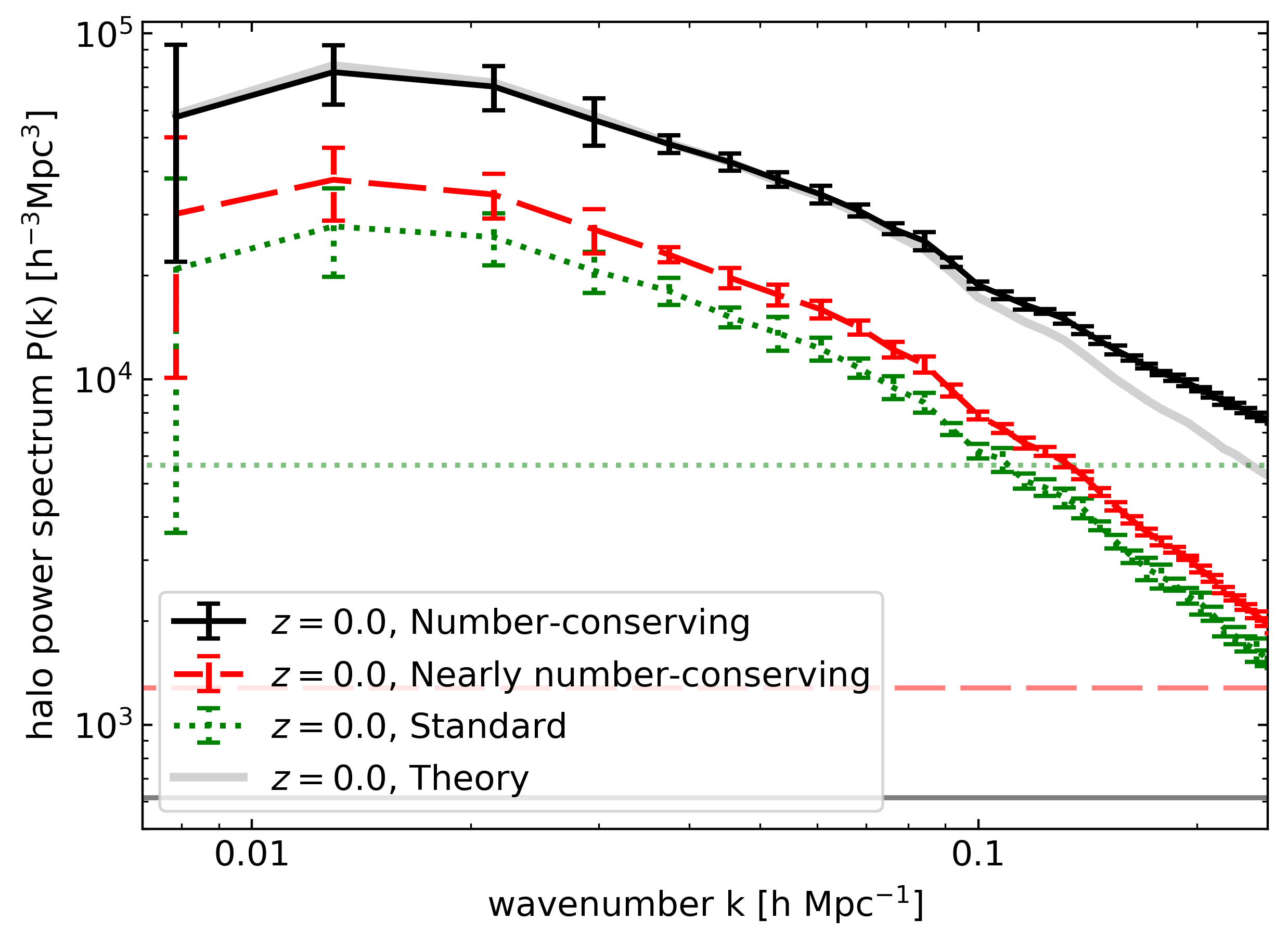}
		\caption{Power spectra at $z=0$ of the halo samples
 shown in the dark gray band in Figure~\ref{fig:haloMassFunction}. Various curves show the power spectra of the halo samples at $z=0$ evolved via the standard (dotted), the number-conserving (solid), and the nearly number-conserving (dashed) merger trees from the halo sample at~$z_\star$. The light gray curve (solid) shows the theoretical
prediction at $z=0$ of the Lagrangian perturbation theory based on the bias parameter of the halos at~$z_\star$. It matches the power spectrum of the halo sample (solid) following the number-conserving merger tree. Horizontal curves show the shot noises.
        }
		\label{fig:haloPowerSpectrum}
		}
\end{figure}

We are now ready to quantify how significant the number conservation is
in the Lagrangian perturbation theory for biased tracers.
It will be addressed with two separate questions: how badly is the number conservation of halos violated? and how well do the Lagrangian perturbation theory predictions for halos match the halo power spectra in simulations?
The implication for its applications in practice will be addressed in
Section~\ref{sec:conclusion}.

First, we investigate how badly the number conservation of halos is violated through mergers, and Figure~\ref{fig:haloMassFunction} shows how halos evolve in time. Consider a halo sample in a narrow mass bin $6\times10^{11} \leq  M (h^{-1}M_{\odot}) < 2\times10^{12}$ at~$z_\star=2.9$ shown in the dark gray band, and the halos in the sample grow in mass via accretion and merger in time. According to the standard merger tree, the distribution of this halo sample at $z=0$ is shown as the dotted curve. The total number of halos in the sample at~$z_\star$ is reduced by~89\% at $z=0$, as the halos in the sample merge with larger halos and they lose their identity. Furthermore, their mass distribution today is broader than the initial mass range at~$z_\star$. The same exercise can be performed backward in time. Consider a halo sample in a narrow mass bin $5\times10^{13} \leq M (h^{-1}M_{\odot}) < 5\times10^{14}$ at $z=0$ 
shown in the light gray band, which can be representative of an observed galaxy sample today. According to the standard merger tree, the distribution of this halo sample at early time~$z_\star$ is shown as the dot-dashed curve. The progenitor halos that make up the light gray band today originate from a very broad distribution in mass at early time (1.1\% of the progenitors at~$z_\star$ are below the mass resolution). Similar trends are observed across halo samples of different mass. This simple exercise confirms the well-established fact that halos are not static, identical objects throughout the evolution like dark matter particles, but evolving, number-conservation-violating objects that change their properties like mass and shape via mergers and accretions. Of course, galaxies with various observable properties follow much more complicated evolutionary tracks.

\begin{figure}[t]
{\centering
\includegraphics[width=\linewidth]{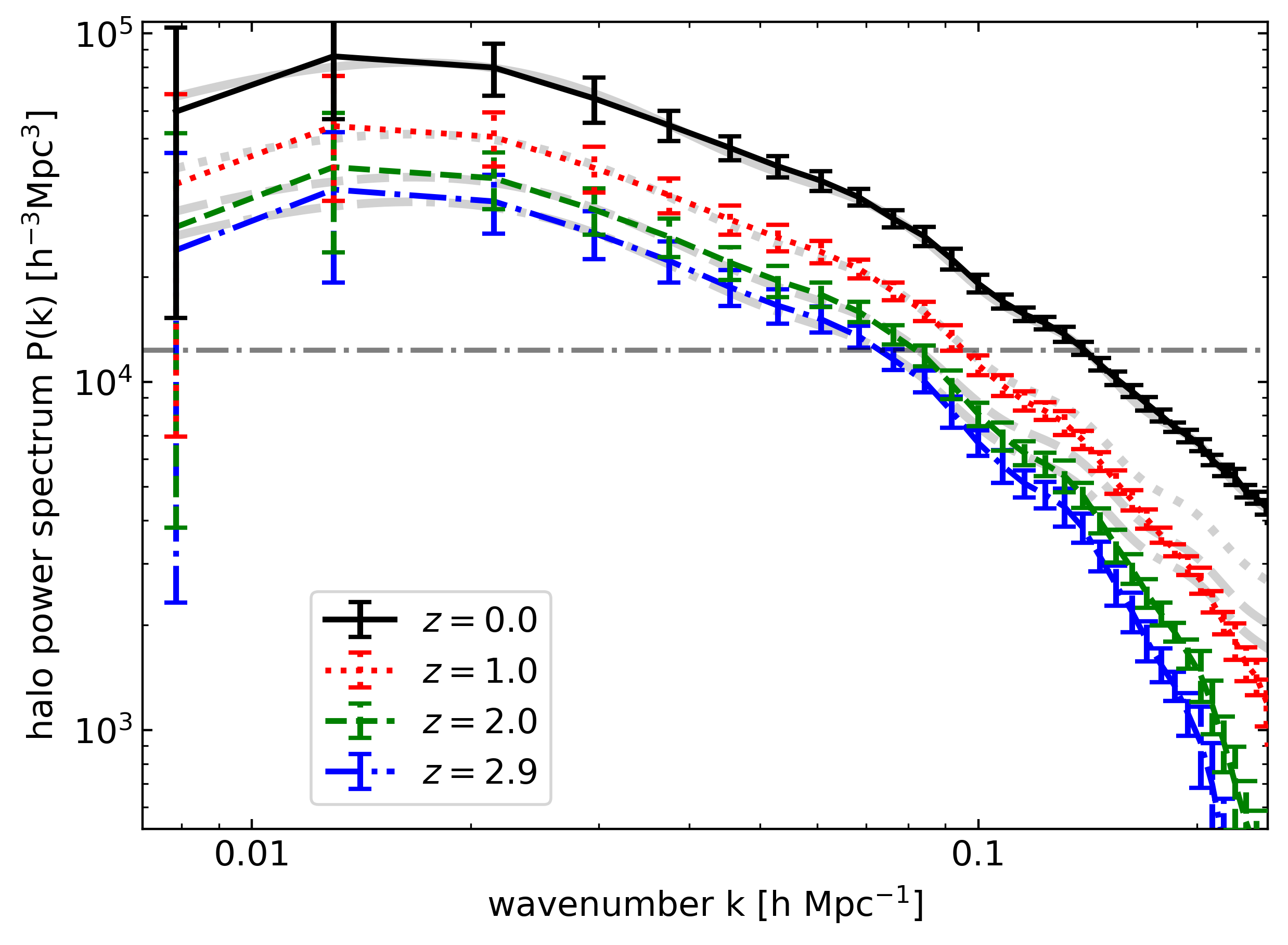}
\caption{Power spectra at various redshifts of the halo sample
 shown in the light gray band in Figure~\ref{fig:haloMassFunction}
at~$z=0$. Various other curves show the power spectra of the halo sample at 
each redshift traced backward in time by considering those particles
that belong to the same halo at~$z=0$ as a distinct halo at~$z>0$ and
using the center-of-mass position as the position of the halo at~$z>0$.
 The gray curves show the theoretical predictions at each redshift
 of the Lagrangian perturbation theory based on the bias parameter 
of the halos at~$z=0$. They are in good agreement with the simulation results.
}\label{fig:back}
}
\end{figure}

To test the significance of the number conservation in the Lagrangian
perturbation theory for biased tracers, we construct a halo sample by 
artificially imposing the number conservation of halos and tracking their evolution from~$z_\star$ to $z=0$. Solid and dashed curves in Figure~\ref{fig:haloMassFunction} illustrate the halo distributions at $z=0$ from the halo sample at~$z_\star$ in the dark gray band, according to the number-conserving (solid) and the nearly number-conserving (dashed) schemes, in which we always keep the 
identities of the progenitor halos, even when they merge into a larger halo. Consequently, two or more halos separately identified at~$z_\star$ could end up in the same halo at~$z=0$. In the number-conserving scheme, those in the same halo are all counted as individual halos at the same position, while in the nearly number-conserving scheme this double counting is lifted (hence the number conservation is slightly violated). 

For both cases (solid and dashed) the halos at~$z_\star$ in the dark gray band are spread over a larger range of mass than in the standard merger tree (dotted). Certainly, many of those halos at~$z_\star$ that disappear in the standard merger tree are part of massive halos at $z=0$. For example, those halos at~$z_\star$ comprise only~6\% of the total mass in the histogram with $M \geq 10^{15} ~h^{-1} M_{\odot}$, but we keep their identities in these two  merger trees. In particular,~52\% of the halos at~$z_\star$ are double counted at $z=0$, shown as the difference between solid and dashed curves. The halo sample in the dark gray band at~$z_\star$ maintains its identities in the solid curve at $z=0$ with the number of halos conserved, and hence its evolution can be correctly modeled by the Lagrangian perturbation theory.

To answer the second question, we measure  in Figure~\ref{fig:haloPowerSpectrum}
the power spectra of three halo samples at $z=0$ evolved from the same halo sample in the dark gray band at~$z_\star$, according to the standard (dotted), the number-conserving (solid), and the nearly number-conserving (dashed) merger trees.
 Compared to the standard merger tree, two other number-conserving trees have more halos in number, or lower shot noises. The distribution over halo mass in Figure~\ref{fig:haloMassFunction} determines the bias of each halo sample, and the presence of more massive halos in the number-conserving merger tree boosts the halo power spectrum. Finally, the light gray curve is the
theoretical
 prediction for the halo power spectrum at $z=0$ from the halo sample in the dark gray band at~$z_\star$, where we used Equation~(\ref{eqn:eulerianBiases}) to predict~$b_1(z=0)$ by measuring~$b_1(z_\star)$ from simulations. The agreement of the theoretical
prediction (gray) with the halo power spectrum (solid)
from the number-conserving merger tree underpins that 
the number conservation is crucial in the Lagrangian perturbation theory for biased tracers. 
Mind that two other power spectra (dotted, dashed) without the number
conservation are in significant disagreement with the theoretical
prediction (gray).

In Figure~\ref{fig:back} we test the significance of the number conservation
in the standard approach to the Lagrangian perturbation theory for biased
tracers. As shown in Figure~\ref{fig:haloMassFunction}, the halo sample
at~$z=0$ originates from the particles that belong to various halos of widely
different mass or from the particles unbound to any halos. Though those
particles cannot be identified as physically distinct halos, we use the
particle identification from the simulations to locate their positions at
high redshifts and assign their center-of-mass positions to the positions
of the halo sample at~$z>0$. Various curves in Figure~\ref{fig:back}
show the power spectra of the halo sample at~$z>0$. In the same way,
we first measure the bias parameter from the halo sample at $z=0$
and use Equation~(\ref{eqn:eulerianBiases})
to predict the bias parameters at high redshifts. 
The agreement of the theoretical predictions (gray) in the Lagrangian
perturbation theory with the simulation results once again confirms
that the number conservation is crucial in the Lagrangian perturbation
theory predictions.
All the Figures are computed by averaging the halo samples from our ten simulations.

\section{Conclusion and Discussion}\label{sec:conclusion}
The Lagrangian perturbation theory (\citealt{Zel'dovich_1970};
see, e.g., \citealt{Buchert_1992, Bouchet_1992, Buchert_1993, Buchert_1994_MNRAS, Bouchet_1995, Catelan_1995, BECOET02}) is a very powerful tool for modeling the nonlinear matter power spectrum due to its simple analytical expression in Equation~(\ref{eqn:LPT_theoryPower}).
Its applications to biased tracers such as dark matter halos and galaxies are widely popular for the same reason (see, e.g., \citealt{Schneider_1995, Taylor_1996, Matsubara_2008_v77}), and the nonlinear power spectrum for biased tracers can be obtained from
\begin{eqnarray}\label{eqn:haloPerturbation}
    \delta_h(\bm k, t) = \int d^3 q ~e^{-i \bm k \cdot (\bm q + \bm \Psi)} \left[1 + \delta_h(\bm q, t_i)\right] ~,
\end{eqnarray}
\noindent for $k>0$.
In this work, we have shown that the number conservation is crucial
for the validity of  this simple analytic expression for biased tracers.
The number conservation is trivially satisfied for the evolution of 
dark matter particles, but biased tracers evolve through mergers and
accretion of continuous matter.

By using numerical simulations and dark matter halos as our biased tracers, we have demonstrated that the Lagrangian perturbation theory prediction overestimates the halo power spectrum at $z=0$ by a factor of~2.8 for a halo sample of mass 
$6\times10^{11} \leq M (h^{-1}M_{\odot}) < 2\times10^{12}$
defined at~$z_\star=2.9$, if we trace their evolution according to the standard
merger tree and hence the number conservation is violated. To demonstrate
the significance of the number conservation,
we have rectified the halo evolution by continuously tracing the individual halos from initial time until present, even if they merge into a larger halo and make up a tiny fraction of the merged halo. For the halo sample constructed this way, the number of halos stays constant throughout the evolution, and not surprisingly the Lagrangian perturbation theory prediction is in good agreement with the numerical simulation output. 

Furthermore, we use the numerical simulations to test the significance
of the number conservation in the standard approach to the Lagrangian
perturbation theory. We chose a halo sample in a  mass bin
 $5\times10^{13} \leq M (h^{-1}M_{\odot}) < 5\times10^{14}$ at $z=0$ to
trace backward in time the dark matter particles that belong to the halo 
sample. The particles in a given halo spread in space and become unbound
to each other, as we move backward in time. 
By treating those particles as a distinct halo at~$z>0$, however,
the number of halos is conserved throughout the backward evolution
in the standard approach. Assigning the center-of-mass position of the particles
to a halo position, we have demonstrated that the Lagrangian perturbation
theory predictions for the halo power spectrum at~$z>0$ match the
numerical simulation results, as implemented in the standard approach
to the Lagrangian perturbation theory.

As opposed to the case in numerical simulations,
the halo number density~$n_h(\bm q,t_i)$ or the fluctuation~$\delta_h(\bm q,
t_i)$ in Equation~\eqref{eqn:haloPerturbation} is in practice
not available in modeling the halo sample in the Eulerian frame.
Real applications of the Lagrangian perturbation theory for biased tracers,
however, involve marginalization of unknown bias parameters,
 rather than taking the bias parameters in the Lagrangian frame as 
input parameters.  Hence the lack of information about those halos
in the Lagrangian frame poses  no real problem. Furthermore,
as demonstrated in our numerical experiment, those bias parameters can be fixed
from simulations as long as the number conservation is correctly imposed
for those halos throughout the evolution.

In the Press-Schechter formalism \citep{PRSC74,BBKS86,BCEK91}, the halo mass
function at high redshift is used to compute the bias parameters
in the Lagrangian space, and the bias parameter in the Eulerian space
is then obtained by using Equations~\eqref{eqn:linearBias} 
and~\eqref{eqn:eulerianBiases}, under the assumption that the halo number
is conserved
\citep{MOWH96,CALUET98,SHTO99,BECOET02,DESJA08,Matsubara_2008_v78,Matsubara_2011,Carlson2013,WHITE14,MADE16,DEJESC18}. 
With halos physically identifiable at high~$z$ as counted in the halo
mass function in the Press-Schechter formalism, their physical
 evolution would be
similar to those in our standard merger tree, as halos evolve through
mergers and accretion, violating the number conservation.
Hence, the predictions for the Eulerian bias parameters made
in the Press-Schechter formalism with the number conservation should then
correspond to the halo samples in our number-conserving merger tree.

Recent modeling of biased tracers in the Lagrangian perturbation theory
includes higher-order bias parameters such as~$b_2$, the tidal bias~$b_{s^2}$,
and so on. These bias parameters are associated with the corresponding
higher-order operators~${\cal O}({\bm q})\ni \delta^2,~s^2$, 
which are often described in the field-level models.
Furthermore, it was recently shown \citep{VLCAWH16,Schmittfull_2019,CHVLWH20,Schmidt_2021} that the field-level modeling of biased tracers in fact works extremely well. With large degree of freedom in the bias parameters, cosmological parameters are successfully extracted \citep{NIDAET20,DAGLET20,IVSIZA20,CHVLET21} after marginalizing over bias parameters. 

In the field-level modeling, 
Equation~\eqref{eqn:haloPerturbation} yields that
the halo number density fluctuation~$\delta_h({\bm q},t_i)$ in
the Eulerian frame is described in terms of operators~$\tilde {\mathcal O}
_i({\bm x},t)$ in the Eulerian frame for each arbitrary bias parameter,
where
\begin{eqnarray}
\label{shifted}
    \tilde {\cal O}_i(\bm k,t) := \int d^3q~{\cal O}_i(\bm q)~e^{-i \bm k \cdot (\bm q+\bm \Psi)}~.
\end{eqnarray}
These are called the {\it shifted}
operators, as the computations in the field-level models 
start with operators~${\cal O}_i({\bm q})$ in the Lagrangian frame 
by shifting them to the Eulerian frame~$\tilde{\cal O}_i({\bm x},t)$ 
according to the displacement field (a.k.a. shifted operators).
Given a series of shifted operators, halos are then modeled with arbitrary bias parameters, which in the end amounts to the effective field theory approach in the Eulerian frame \citep{Mirbabayi_2015, Senatore_2015, VLWHAV15,IVSIZA20} (but of course with different operators).

However, without the number 
conservation of any field associated with each operator~${\cal O}_i$, 
the shifted operator in Equation~\eqref{shifted}
is not the Fourier transform of the field evolved from ${\cal O}_i(\bm{q})$ 
in the Lagrangian frame. In other words, this procedure of 
shifting an operator ${\cal O}_i(\bm{q})$ in the Lagrangian space to Eulerian
space by using $\bm{x}=\bm{q}+\bm{\Psi}$ is not equivalent to
Equation~\eqref{shifted}.
The notation here is again a misnomer in a sense that $n_h(\bm{q},t_i)$
represents not physical 
halos at~$t_i$, but the particle distribution at~$t_i$
that eventually form halos in the Eulerian frame.
Hence, this procedure works just fine, because in the end 
$\tilde {\cal O}_i(\bm{k},t)$ is some operator in the Eulerian space defined in terms of~${\mathcal O}_i(\bm q)$
in the Lagrangian space
with large-scale bulk motion accounted for. But it is clear that two fields~${\cal O}_i(\bm{q})$ and~${\cal O}_i(\bm{x},t)$ are not related.

\section*{Acknowledgments}
We acknowledge useful discussions with Julian Adamek, Andrej Obuljen, and Zvonimir Vlah. 
This work is supported by the Swiss National Science Foundation Grant CRSII5\_198674. We acknowledge access to Piz Daint and Alps at the Swiss National Supercomputing Centre under the project ID UZH-24.

\bibliographystyle{aasjournal.bst}
\bibliography{LPT4halos.bib}

\end{document}